\documentclass[preprint,aps,showpacs]{revtex4}
\usepackage{amsfonts}
\usepackage{amsmath}
\usepackage{amssymb}
\usepackage{graphicx}
\usepackage{epsfig}
\begin{document}
\title{Symmetric Device-Independent Quantum Key Distribution Against General Attack}
\author{Yong-gang Tan}
\email{yonggang.tan@gmail.com} \affiliation{Physics and Information
Engineering Department, Luoyang Normal College, Luoyang 471022,
Henan, People's Republic of China}

\begin{abstract}

A symmetric device-independent quantum key distribution (DIQKD)
protocol is proposed in this paper, with Holevo limit and
subadditivity of von Neumann entropy, one can bound Eve's ability
with collective attack. Together with symmetry of this protocol, the
state Eve prepared for Alice and Bob, and at the same time, her
eavesdropping on Alice's and Bob's measurements can be definitely
inferred at the assumption that Eve aims at maximizing her
information gain. The optimal state under this circumstance can be
solely bounded with Alice and Bob's statistical results on the
quantity of Clauser-Horne-Shimony-Holt (CHSH) polynomial $S$, that
is, our symmetric DIQKD has the same secure basis as that of Ekert91
protocol.



\end{abstract}
\maketitle

\section{introduction}

Quantum key distribution (QKD) is an art of generating physically
secure key between remote partners, the information sender Alice,
and the receiver Bob~\cite{bb84,epr91,gsin02}, even if in the
presence of a powerful eavesdropper, namely Eve, whose capability is
only limited by quantum mechanics. On one hand, the security proof
of QKD has been obtained with nearly perfect
apparatus~\cite{lc99,sp00}. On the other hand, there are different
loopholes in current QKD experiments that may injure the security of
the final key
bits~\cite{higm95,qflm07,zfqcl08,fqtl07,xql10,lwwesm10,gllskm11}.
Even with the perfect experimental apparatus~\cite{lc99,sp00}, there
are also some self-evident assumptions that guarantee the security
of final key bits. For instance, we have to assume that Alice and
Bob have the freedom to choose the bases for their preparations and
measurements. Their classical results which is unwanted to be leaked
out should be completely secret. At the same time, Alice and Bob
should entirely control their apparatus to generate the raw keys. Or
else, the final key bits cannot be secure.

As for commercial application, the apparatuses of Alice and Bob will
be black boxes that may be provided by their potential rivals. It is
interesting how Alice and Bob can determine the security of their
final key bits extracted from these black boxes? Recently, the
device-independent QKD (DIQKD)~\cite{my98,abgmps07,pabgms09} has
been suggested to ask for the answer. It was assumed in this
protocol that Alice and Bob have no knowledge about their
measurement devices. The violation of CHSH inequality will impose
restriction on Hilbert space dimension of their measurements to
ensure the efficient quantum correlations between Alice and
Bob~\cite{chsh69,bell65}. Secure key bits against collective attack
for this protocol has been proven~\cite{abgmps07,pabgms09}. Its
final key generation rate depends on two parameters, the quantity of
CHSH polynomial $S$ and the quantum bit error rate (QBER) $Q$. These
two parameters are decided by the state measured by the legitimate
users' devices and the way of their measurements at the same time.
As Alice and Bob have no idea about the state prepared by Eve, and
their measurement devices can also be fabricated by their rivals,
generalization from collective attack to general attack is still
missed.

The way of state preparation in DIQKD is the same as those of
Ekert91 protocol~\cite{epr91} and entanglement-based QKD protocol
with sources in the middle~\cite{ma07,renner05a} where Eve's
eavesdropping ability is bounded with collective attack as quantum
De Finetti theorem can be applied after Alice and Bob having
randomized the measurement sequences on their
states~\cite{renner05,renner07}. In DIQKD, however, It is impossible
for Alice and Bob to make sure that their measurements function
exactly on the quantum systems as their expectations. In fact, Eve
may devise Alice's and Bob's measurements differently in every run.
In this paper, a symmetric DIQKD protocol is proposed. The symmetry
of this protocol, together with the Holevo
limit~\cite{holevo73,nielsen00,cabello00}, will provide strong
confinements on Eve's eavesdropping. We show Eve's information is
maximized when all states distributed to Alice and Bob are
identically prepared. Then the procedure of uniform their states is
completed automatically, and Alice and Bob can estimate their
parameters by randomizing the sequences of their classical results.
Furthermore, Eve's optimal state when her illegal information is
maximized can be solely bounded with Alice and Bob's parameter $S$.
Then our symmetric DIQKD has the same secure basis as that of
Ekert91 protocol.

\section{A symmetric DIQKD protocol}

Our DIQKD protocol is symmetric not only because Alice's and Bob's
basis choices are symmetric, but also because the statistical
results generated from all bases are the same. It works as follows.
(1) $N$ EPR pairs emit from the signal source set between Alice's
and Bob's labs. One particle of the EPR pair is sent to Alice and
the other one is sent to Bob. (2) Both Alice and Bob choose four
expecting measurement bases as $\theta_{1}=\sigma_{x}$,
$\theta_{2}=(\sigma_{x}+\sigma_{z})/\sqrt{2}$,
$\theta_{3}=\sigma_{z}$,
$\theta_{4}=(-\sigma_{x}+\sigma_{z})/\sqrt{2}$ (As is shown in Fig.
1). In each run, Alice will randomly measure the incoming particle
in one of the four bases, and so does Bob. (3) After all EPR pairs
having been distributed, Alice and Bob announce the bases they used
in each run through their classical channels. (4) Alice and Bob
randomize the sequences of their classical results. They keep
partial measurement results on the same bases as secrecy that will
be used to generate secure final key bits. Then they publish all the
other measurement results to estimate the disturbances and
correlations on their sifted key bits. They abort their
communication if the parameter estimation fails to meet their
predefined requirements. Or else, they carry out privacy
amplification to generate their secure final key.
\begin{figure}[!ht]
\centering
\includegraphics[width=1\textwidth]{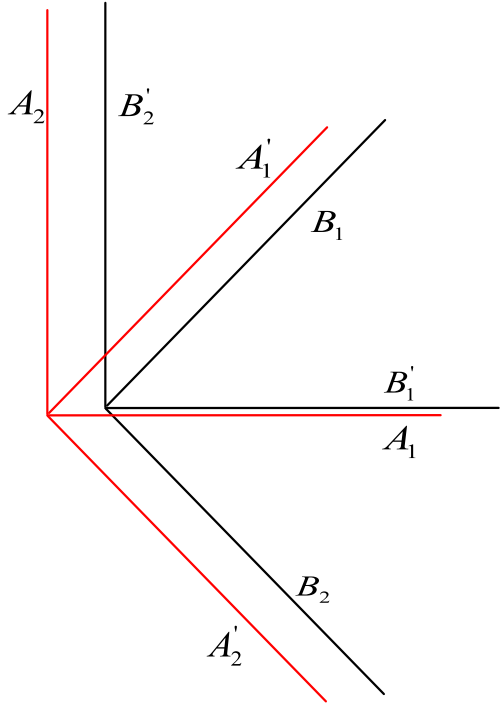}
\caption{Schematic for symmetric DIQKD. $A_{i}$s and
$A_{i}^{\prime}$s are Alice's possible basis choices. $B_{j}$s and
$B_{j}^{\prime}$s are Bob's basis choices.}
\label{fig:graph1}
\end{figure}

Without loss of generality, we assume Eve prepares all Alice's and
Bob's systems and her auxiliary systems in a big state
$\rho_{A_{1\cdots{N}}B_{1\cdots{N}}E}$. Noting Alice's measurement
operations as $A_{1}$, $A_{2}$, $A_{1}^{\prime}$ and
$A_{2}^{\prime}$, Bob's measurement operations as $B_{1}$, $B_{2}$,
$B_{1}^{\prime}$ and $B_{2}^{\prime}$, the joint measurements of
Alice's and Bob's devices can then be depicted as
$A^{u_{1}}_{i_{1}}(t_{1})\otimes{B^{v_{1}}_{j_{1}}(t_{1})}\cdots{A^{u_{N}}_{i_{N}}(t_{N})}\otimes{B^{v_{N}}_{i_{N}}(t_{N})}$.
Here $t_{1}$, $\cdots$, ${t_{N}}$ are sorted in time sequence as Eve
may eavesdrop on Alice's and Bob's measurements differently in every
run, $u_{1}$, $\cdots$, $u_{N}$, $v_{1}$, $\cdots$, $v_{N}$
correspond to the upper indexes, and $i_{1}$, $\cdots$, $i_{N}$ can
be $1$ or $2$ randomly. As Eve's measurements will not affect the
marginal distributions of Alice's and Bob's classical results, their
results generated from $\rho_{A_{1\cdots{N}}B_{1\cdots{N}}E}$ can be
depicted as
$Tr[A^{u_{1}}_{i_{1}}(t_{1})\otimes{B^{v_{1}}_{j_{1}}(t_{1})}\cdots{A^{u_{N}}_{i_{N}}(t_{N})}\otimes{B^{v_{N}}_{i_{N}}(t_{N})}Tr_{E}(\rho_{A_{1\cdots{N}}B_{1\cdots{N}}E})]$,
where $Tr_{E}$ is the trace on Eve's systems. According to DIQKD
protocol, the result in the $k$th run is
$Tr(A^{u_{k}}_{i_{k}}(t_{k})\otimes{B^{v_{k}}_{i_{k}}}(t_{k})(Tr_{A_{1\cdots{k-1}k+1\cdots{N}}B_{1\cdots{k-1}k+1\cdots{N}}E}(\rho_{A_{1\cdots{N}}B_{1\cdots{N}}E}))=a^{k}b^{k}$,
where $a_{k}$ and $b_{k}$ are the classical results obtained by
Alice and Bob respectively. If they are binary, taking $-1$ and $1$
for example, it is proven that $A^{u_{k}}_{i_{k}}(t_{k})$ and
$B^{v_{k}}_{i_{k}}(t_{k})$ functioned on qubit
states~\cite{abgmps07,pabgms09,tsirelson80,tsirelson93,masanes06}.
Furthermore, $A_{1_{k}}(t_{k})$, $A_{2_{k}}(t_{k})$,
$B_{1_{k}}(t_{k})$ and $B_{2_{k}}(t_{k})$ can be set in the same
plane $x-z$. Similarly, $A^{\prime}_{1_{k}}(t_{k})$,
$A^{\prime}_{2_{k}}(t_{k})$, $B^{\prime}_{1_{k}}(t_{k})$ and
$B^{\prime}_{2_{k}}(t_{k})$ can be set in another plane
$x^{\prime}-z^{\prime}$~\cite{abgmps07,pabgms09}.

Defining
$Tr_{A_{1\cdots{k-1}k+1\cdots{N}}B_{1\cdots{k-1}k+1\cdots{N}}E}(\rho_{A_{1\cdots{N}}B_{1\cdots{N}}E})\equiv\rho^{k}_{AB}$,
the Hilbert space of Alice and Bob as $H_{AB}$,
$\rho^{k}_{AB}\in{H_{AB}}$ and dim$H_{AB}\le4$ should be satisfied.
According to Holevo limit~\cite{holevo73,cabello00,nielsen00}, Eve's
ability to distinguish the state shared by Alice and Bob is limited
by $S(Tr_{E}(\rho_{A_{1\cdots{N}}B_{1\cdots{N}}E}))$, where
$S(\rho)=-Tr(\rho\log_{2}(\rho))$ is the von Neumann entropy. If
writing $\rho_{A_{1\cdots{N}}B_{1\cdots{N}}E}$ as
$\rho_{A_{1\cdots{s}s+1\cdots{N}}B_{1\cdots{s}s+1\cdots{N}}E}$,
$S(Tr_{E}(\rho_{A_{1\cdots{N}}B_{1\cdots{N}}E}))\le{S(Tr_{A_{s+1\cdots{N}}B_{s+1\cdots{N}}E}(\rho_{A_{1\cdots{N}}B_{1\cdots{N}}E}))}+S(Tr_{A_{1\cdots{s}}B_{1\cdots{s}}E}(\rho_{A_{1\cdots{N}}B_{1\cdots{N}}E}))$
is required for the subadditivity of entropy. With the same
procedure, one can have
$S(Tr_{E}(\rho_{A_{1\cdots{N}}B_{1\cdots{N}}E}))\le\sum_{k}S(Tr_{A_{1\cdots{k-1}k+1\cdots{N}}B_{1\cdots{k-1}k+1\cdots{N}}E}(\rho_{A_{1\cdots{N}}B_{1\cdots{N}}E}))$.
The equality holds if and only if
$Tr_{E}(\rho_{A_{1\cdots{N}}B_{1\cdots{N}}E})$ can be written as $N$
product systems shared between Alice and Bob, that is,
$Tr_{E}(\rho_{A_{1\cdots{N}}B_{1\cdots{N}}E})=\otimes_{k=1}^{N}\rho^{k}_{AB}$.

Eve controls the transmission of quantum state, thus she can make
$\rho^{k}_{AB}$ optimal for her information gain. As
dim$H_{AB}\le4$, projective measurements can be launched on
$\rho^{k}_{AB}$ in a 4-dimensional Hilbert space $H$, with
$H_{AB}\subseteq{H}$. Let $\sum_{l}P^{l}_{AB}=I$ be projective
measurements in $H$, it is proven that
$S(\sum_{l}P^{l}\rho^{k}_{AB}P^{l})\ge{S(\rho^{k}_{AB})}$. The
equality holds if and only if
$\rho^{k}_{AB}=\sum_{l}P^{l}\rho^{k}_{AB}P^{l}$~\cite{nielsen00}.
That is, $\sum_{l}P^{l}\rho^{k}_{AB}P^{l}$ can be diagonalized in
bases $T=\{\tau_{l}\}$ with $\tau_{l}\tau^{\dag}_{l}=P^{l}$.
Defining
$\Lambda\equiv\sum_{l}P^{l}\rho^{k}_{AB}P^{l}=P^{-1}\rho^{k}_{AB}P$,
then $\Lambda$ is a diagonal matrix in $H$, and $P$ is composed with
the basis vectors $\{\tau^{k}\}$. Noticing Bell bases
$B=\{\varsigma_{l}\}$ is also a set of bases in $H$, there should be
a unitary operator $U$ satisfying $T=UB$. Then one can have
$\Lambda=B^{-1}U^{-1}\rho^{k}_{AB}UB$, which means $\rho^{k}_{AB}$
can be diagonalized in Bell bases after it has been operated as
$U^{-1}\rho^{k}_{AB}U$. It is apparently that this process will not
alter the amount of entanglement on $\rho^{k}_{AB}$, and Eve's
information gain is only determined by the elements of $\Lambda$.
That is, the assumption that the state $\rho^{k}_{AB}$ can be
diagonlized on Bell bases will not affect Eve's information gain,
and at the same time, it will not harm Eve's ability to intervene
Alice and Bob's communication. Then it does not loss any generality
to assume $\rho^{k}_{AB}$ can be diagonalized on Bell bases so long
as both Alice's and Bob's marginal distributions are
symmetric~\cite{abgmps07,pabgms09}.

Suppose the state distributed by Eve in the $k$th run is
$\sigma^{k}_{AB}\equiv(1-p_{k})\rho_{|\Phi^{+}\rangle}+p_{k_{1}}\rho_{|\Phi^{-}\rangle}+p_{k_{2}}\rho_{|\Psi^{+}\rangle}+p_{k_{3}}\rho_{|\Psi^{-}\rangle}$,
where $p_{k}=p_{k_{1}}+p_{k_{2}}+p_{k_{3}}$,
$|\Phi^{+}\rangle=\frac{1}{\sqrt{2}}(|00\rangle+|11\rangle)$,
$|\Phi^{-}\rangle=\frac{1}{\sqrt{2}}(|00\rangle-|11\rangle)$,
$|\Psi^{+}\rangle=\frac{1}{\sqrt{2}}(|01\rangle+|10\rangle)$, and
$|\Psi^{-}\rangle=\frac{1}{\sqrt{2}}(|01\rangle-|10\rangle)$. We now
show that Eve's information on Alice and Bob's results is maximized
when all $\sigma^{k}_{AB}$s are identically prepared.

(I) When there are only two types of quantum states equiprobably
prepared on Alice and Bob's $N$ shared systems (This assumption does
not loss any generality if these two states are prepared to be the
same), they can be denoted as
$\sigma^{(\alpha)}_{AB}=(1-p^{(\alpha)})\rho_{|\Phi^{+}\rangle}+p^{(\alpha)}_{1}\rho_{|\Phi^{-}\rangle}+p^{(\alpha)}_{2}\rho_{|\Psi^{+}\rangle}+p^{(\alpha)}_{3}\rho_{|\Psi^{-}\rangle}$
and
$\sigma^{(\beta)}_{AB}=(1-p^{(\beta)})\rho_{|\Phi^{+}\rangle}+p^{(\beta)}_{1}\rho_{|\Phi^{-}\rangle}+p^{(\beta)}_{2}\rho_{|\Psi^{+}\rangle}+p^{(\beta)}_{3}\rho_{|\Psi^{-}\rangle}$.
Then Eve's information gain on these states should be represented as
$\frac{N}{2}S(\sigma^{(\alpha)}_{AB})+\frac{N}{2}S(\sigma^{(\beta)}_{AB})$.
If the statistical state on the $N$ entangled systems measured by
Alice and Bob is
$\sigma_{AB}=(1-p)\rho_{|\Phi^{+}\rangle}+p_{1}\rho_{|\Phi^{-}\rangle}+p_{2}\rho_{|\Psi^{+}\rangle}+p_{3}\rho_{|\Psi^{-}\rangle}$,
we have $p^{(\alpha)}_{1}=p^{(\beta)}_{1}=p_{1}$,
$p^{(\alpha)}_{2}=p^{(\beta)}_{2}=p_{2}$, and
$p^{(\alpha)}_{3}=p^{(\beta)}_{3}=p_{3}$ when Eve's information gain
is optimal.

(II) When there are $m$ types of density matrixes equiprobably
prepared by Eve, we assume Eve's illegal information is maximal only
when all of them are identically prepared on the $N$ entangled
systems between Alice and Bob. That is,
$S(\sigma^{(\alpha)}_{AB})+S(\sigma^{(\beta)}_{AB})+\cdots+S(\sigma^{(m)}_{AB})\le{m}S(\sigma_{AB})$
is satisfied with the equality holds if and only if
$\sigma^{(\alpha)}_{AB}=\sigma^{(\beta)}_{AB}=\cdots=\sigma^{(m)}_{AB}=\sigma_{AB}$.
Here $\sigma_{AB}$ is the statistical expression on the $N$ shared
systems of Alice and Bob.

(III) When there are $m+1$ types of states measured by Alice and Bob
equiprobably, Eve's information gain on them can be written as
$S_{E}=\frac{N}{m+1}[S(\sigma^{(\alpha)}_{AB})+S(\sigma^{(\beta)}_{AB})+\cdots+S(\sigma^{(m)}_{AB})+S(\sigma^{(m+1)}_{AB})]$.
If the statistical expression for the first $m$ types of states on
the $\frac{m}{m+1}N$ systems is
$\sigma^{(\Omega)}_{AB}=(1-p^{(\Omega)})\rho_{|\Phi^{+}\rangle}+p^{(\Omega)}_{1}\rho_{|\Phi^{-}\rangle}+p^{(\Omega)}_{2}\rho_{|\Psi^{+}\rangle}+p^{(\Omega)}_{3}\rho_{|\Psi^{-}\rangle}$,
$S_{E}$ can be bounded as
$\frac{N}{m+1}[mS(\sigma^{(\Omega)}_{AB})+S(\sigma^{(m+1)}_{AB})]$
according to step (II). With simple calculation, one can obtain its
maximum value when
$\sigma^{(\Omega)}_{AB}=\sigma^{(m+1)}_{AB}=\sigma_{AB}$, with
$\sigma_{AB}$ being the statistical representation of all $N$ shared
systems between Alice and Bob. In conclusion, Eve should prepare the
state on the $N$ systems identically if she want to maximize her
illegal information from Alice and Bob's communication. Then there
is no need to randomize the measurement sequences to make their
states uniformly distributed. Quantum De Finetti theorem can be
applied in DIQKD protocol and Eve's information can be bounded with
collective attack~\cite{renner05,renner07}.

Alice and Bob may not infer Eve's single intervention on their
measurements in the $k$th run because of quantum randomness. As
their states are identically prepared, however, they can deduce the
equivalent operations averaged from all results. Denoting the
equivalent operations in $x-z$ plane as $A_{1}$, $A_{2}$, $B_{1}$
and $B_{2}$, their directions are assumed to be $\theta_{1}$,
$\theta_{2}$, $\varphi_{1}$ and $\varphi_{2}$ respectively.
Similarly, denoting the equivalent operations in
$x^{\prime}-z^{\prime}$ plane as $A^{\prime}_{1}$, $A^{\prime}_{2}$,
$B^{\prime}_{1}$ and $B^{\prime}_{2}$, the corresponding directions
for them are $\theta^{\prime}_{1}$, $\theta^{\prime}_{2}$,
$\varphi^{\prime}_{1}$ and $\varphi^{\prime}_{2}$. The CHSH
polynomial can be calculated as
$S=\sum_{i,j}{sign(3.5-i-j)}Tr(A_{i}B_{j}\sigma_{AB})$, and
$S^{\prime}=\sum_{i,j}{sign(3.5-i-j)}Tr(A^{\prime}_{i}B^{\prime}_{j}\sigma_{AB})$,
with $i$, $j=1$, or $2$, and $sign(x)$ getting the sign of $x$. Our
symmetric DIQKD protocol requires $S=S^{\prime}$, moreover, it
requires that all $sign(3.5-i-j)Tr(A_{i}B_{j}\sigma_{AB})$s and
$sign(3.5-i-j)Tr(A^{\prime}_{i}B^{\prime}_{j}\sigma_{AB})$s have the
same value $\frac{S}{4}$. Then one can obtain $p_{1}=p_{2}$, and
$\theta_{2}-\theta_{1}=\frac{\pi}{2}$,
$\varphi_{2}-\varphi_{1}=-\frac{\pi}{2}$,
$\theta_{2}-\varphi_{1}=\frac{\pi}{4}$ and
$\theta_{1}-\varphi_{2}=\frac{\pi}{4}$, or
$\theta_{2}-\theta_{1}=-\frac{\pi}{2}$,
$\varphi_{2}-\varphi_{1}=\frac{\pi}{2}$,
$\theta_{2}-\varphi_{1}=-\frac{\pi}{4}$ and
$\theta_{1}-\varphi_{2}=-\frac{\pi}{4}$. With the same procedure, we
can have the relationships among $\theta^{\prime}_{1}$,
$\theta^{\prime}_{2}$, $\varphi^{\prime}_{1}$ and
$\varphi^{\prime}_{2}$. It is interesting to find that measurements
satisfying the above relationships can be proven to maximize the
value of $S$. For Alice and Bob, bigger $S$ means less disturbances,
then Eve should have the measurements of Alice and Bob in every run
obeys the above relationships in order to conceal her existence.

Two important things may be reconsidered: what is the relationship
between plane $x-z$ and $x^{\prime}-z^{\prime}$ and whether should
Eve prepares the states for $A_{1}$, $A_{2}$, $B_{1}$, $B_{2}$ and
$A_{1}^{\prime}$, $A_{2}^{\prime}$, $B_{1}^{\prime}$,
$B_{2}^{\prime}$ with different systems? It is interesting to notice
that both questions deal with the relationship between measurements
and information. And they can be answered at the same time. If plane
$x-z$ does not coincide with plane $x^{\prime}-z^{\prime}$, they
belong to different Hilbert spaces. When Eve prepares the states on
the same systems, generalized measurements are carried out
inevitably when Eve distributes them in two different Hilbert
spaces. This process will decrease Eve's information gain on the
state~\cite{nielsen00}. However, when Eve prepares the states on
plane $x-z$ and $x^{\prime}-z^{\prime}$ with different systems,
monogamy of entanglement means that Alice's and Bob's results
extracted on the same bases should be totally
uncorrelated~\cite{brub99,ckw00,kw04}. This will increase QBER on
the key, that is, she will risk to be detected on line without
gaining more information. Then for the sake of Eve's optimal
information gain, and concealing her existence at the same time, the
Hilbert space of $x-z$ coincides with that of
$x^{\prime}-z^{\prime}$, and the states for them are prepared on the
same systems correspondingly.

When $p_{1}=p_{2}$, one can have the relationship
$p_{1}+p_{3}=\frac{1}{2}-\frac{S}{4\sqrt{2}}$. If the measurement
directions between Alice and Bob are well aligned, the QBER can be
calculated as $p_{1}+p_{3}$. Or else, it should be written as
$Q=\frac{1-(1-p-p_{3})\cos\vartheta}{2}$, where $\vartheta$ is the
included angle between these measurement directions. This value is
greater than that of the former. Then Alice's measurement bases
should keep alignment with those of Bob, one can obtain
$S=2\sqrt{2}[1-2Q]$, which is the same as that
in~\cite{abgmps07,pabgms09}. In practical implementation of DIQKD,
Alice and Bob can not obtain the value of $p$. Defining
$q\equiv{p+p_{3}}$, we have $q=1-\frac{S}{2\sqrt{2}}=2Q$. But the
exact value of $p_{1}$ and $p_{3}$ is still unknown. As Eve's
information on $\sigma_{AB}$ can be written as
$S(\sigma_{AB})=-(1-q+p_{3})\log_{2}(1-q+p_{3})-2p_{1}\log_{2}p_{1}-p_{3}\log_{2}p_{3}$,
however, we have $p_{3}=\frac{q^{2}}{4}$ and
$p_{1}=\frac{q}{2}-\frac{q^{2}}{4}$ when $S(\rho_{AB})$ is maximal.
That is, the optimal state for Eve's eavesdropping is
$\sigma^{optimal}_{AB}=(1-q+\frac{q^{2}}{4})\rho_{|\Phi^{+}\rangle}+(\frac{q}{2}-\frac{q^{2}}{4})\rho_{|\Phi^{-}\rangle}+(\frac{q}{2}-\frac{q^{2}}{4})\rho_{|\Psi^{+}\rangle}+\frac{q^{2}}{4}\rho_{|\Psi^{-}\rangle}$.
Different to the cases where Alice and Bob having full control of
their measurement devices~\cite{eberhard93,brunner07,cabello07},
color noise is optimal for Eve in the DIQKD protocol. This is
because Alice and Bob can calculate the value of $p$ accurately in
the former but they can only estimate the value of $q$ in the
latter.

Until to now, we have proven Eve's optimal information to be
$S(\sigma^{optimal}_{AB})$. If its corresponding quantity of CHSH
polynomial is $S^{optimal}$, however, can Eve's optimal information
be bounded as $S(\sigma^{optimal}_{AB})$ when Alice and Bob's
statistical value of $S$ is equal to $S^{optimal}$? If not so, there
must be another $\sigma^{optimal\prime}_{AB}$ with which Eve can
obtain more illegal information. That is,
$S(\sigma^{optimal\prime}_{AB})>S(\sigma^{optimal}_{AB})$ is
satisfied. According to the discussion above,
$\sigma^{optimal\prime}_{AB}$ should also be represented as
$\sigma^{optimal\prime}_{AB}=(1-q^{\prime}+\frac{q^{\prime2}}{4})\rho_{|\Phi^{+}\rangle}+(\frac{q^{\prime}}{2}-\frac{q^{\prime2}}{4})\rho_{|\Phi^{-}\rangle}+(\frac{q^{\prime}}{2}-\frac{q^{\prime2}}{4})\rho_{|\Psi^{+}\rangle}+\frac{q^{\prime2}}{4}\rho_{|\Psi^{-}\rangle}$.
Its corresponding $S$ can be calculated to be less than
$2\sqrt{2}(1-q^{\prime})$. As
$S(\sigma^{optimal\prime}_{AB})>S(\sigma^{optimal}_{AB})$, one can
have ${S^{optimal}}\le2\sqrt{2}(1-q^{\prime})\le2\sqrt{2}(1-q)$. For
Alice and Bob, great $S$ means less information can be obtained Eve,
which means $S^{optimal}$ can bound Eve's illegal information.
Generally, if the value of CHSH polynomial is $S$, Eve's illegal
information should be less than
$E(S)=-\frac{1}{4}(1+\frac{S}{2\sqrt{2}})\log_{2}\frac{1}{4}(1+\frac{S}{2\sqrt{2}})-\frac{1}{4}(1-\frac{S}{2\sqrt{2}})\log_{2}\frac{1}{4}(1-\frac{S}{2\sqrt{2}})-(\frac{1}{2}-\frac{S^{2}}{16})\log_{2}(\frac{1}{4}-\frac{S^{2}}{32})$.
Thus, DIQKD can be bounded with the quantity of CHSH polynomial,
which means it has the same secure basis as that of Ekert91
protocol~\cite{epr91}. For collective attack, Alice and Bob's key
generation can be represented as $r=1-H_{2}(Q)-\chi$, where
$H_{2}(Q)=-Q\log_{2}Q-(1-Q)\log_{2}(1-Q)$ is the Shannon entropy and
$\chi$ is the Holevo limit~\cite{biham97,biham97a}. In our symmetric
DIQKD, the lower bound of Alice and Bob's key generation rate can be
estimated as $r\ge1-H_{2}(Q)-S(\sigma^{optimal}_{AB})$. If the
relationship $S=2\sqrt{2}(1-2Q)$ is satisfied, the key rate of our
DIQKD can be calculated as
\begin{equation}
r\ge1-{H_{2}(\frac{1}{2}-\frac{S}{4\sqrt{2}})}-E(S).
\end{equation}

\section{discussion and conclusion}

In this paper, a symmetric DIQKD has been proposed, where Eve's
ability of eavesdropping can be bounded with collective attack. That
is, generalization on the security of DIQKD from collective attack
to general attack can be realized. Its security can be estimated
similarly as that of Ekert91 protocol, that is, determined by the
quantity of CHSH polynomial $S$. However, we have only considered an
ideal case where the loss in the quantum channel is not added in. In
practical implementation of this protocol, there may be detecting
loophole because of imperfectly detecting
efficiency~\cite{brunner07,cabello07}. Especially, faking state
attack has been proposed to eavesdrop on DIQKD protocols with
inefficient measurement devices~\cite{gerhardt11}. To make DIQKD
more practically with present devices, however, proposition for
experimental realization of this protocol has been
given~\cite{gisin10}. Besides its attractiveness of practical
application, DIQKD is physically interesting as it provides us a way
to understand the nonlocality of quantum principles, with which the
legitimate users can set up secure communication without any
knowledge about their quantum objects. The author thank helpful
discussion from Q.-Y. Cai and X. Ma. This work is sponsored by the
National Natural Science Foundation of China (Grant No 10905028) and
HASTIT.

\appendix

\section{Bounding Eve's information with Holevo limit}

In DIQKD, Alice and Bob measure on the state $\rho$ prepared by Eve.
According to the Holevo limit, Eve's illegal information on the
results of Alice and Bob can be bounded with $S(\rho)$. If $\rho$ is
composed with many subsystems, that is, $\rho=\rho_{1,2,\cdots,n}$.
$S(\rho)$ can be proven to satisfy the relationship
\begin{equation}
S(\rho)\le\sum_{i=1}^{n}S(\rho_{i}),
\end{equation}
where $\rho_{i}$ is the state functions with Alice and Bob's
measurements in the $i$th run. This conclusion can easily be proven
with Klein's inequality $S(\rho)\le{-Tr(\rho\log_{2}\sigma)}$.
Defining $\rho\equiv\rho_{1,2,\cdots,n}$,
$\sigma\equiv\otimes_{i=1}^{n}\rho_{i}$, and substituting them into
the Klein's inequality, we have
\begin{equation}
\begin{array}{lll}
S(\rho_{1,2,\cdots,n})&\le&-Tr(\rho_{1,2,\cdots,n}\log_{2}\otimes_{i=1}^{n}\rho_{i})\\
 &=&-Tr(\rho_{1,2,\cdots,n}(\log_{2}\prod_{i=1}^{n}\rho_{i}))\\
 &=&-Tr(\rho_{1,2,\cdots,n}(\sum_{i=1}^{n}\log_{2}\rho_{i}))\\
 &=&\sum_{i=1}^{n}S(\rho_{i}).
\end{array}
\end{equation}
The equality holds if and only if the state can be written as
product states of $n$ systems.

\section{Diagonalizing $\rho_{AB}$ on Bell bases}

Suppose $P_{i}$ is a complete set of orthogonal projectors and
$\rho$ is a density operator. Then the entropy of the state
$\sigma\equiv\sum_{i}P_{i}\rho{P_{i}}$ of the system after the
measurement is at least as greater as the original entropy
\begin{equation}
S(\sigma)\ge{S(\rho)}).
\end{equation}
This results can be verified easily with Klein's inequality.
\begin{equation}
\begin{array}{lll}
S(\rho)&\le&{-Tr(\rho\log_{2}\sigma)}\\
&=&-Tr(\sum_{i}P_{i}\rho\log_{2}\sigma)\\
&=&-Tr(\sum_{i}P_{i}P_{i}\rho\log_{2}\sigma)\\
&=&-Tr(\sum_{i}P_{i}\rho\log_{2}\sigma{P_{i}})\\
&=&-Tr(\sum_{i}P_{i}\rho{P_{i}}\log_{2}\sigma)\\
&=&S(\sigma).
\end{array}
\end{equation}
If $P_{i}$s are the set of projective operators which can maximize
Eve's information after it has functioned on state $\rho$. Writing
$P_{i}$ as $\tau_{i}\tau_{i}^{\dag}$, then $T=\{\tau_{i}\}$ is the
basis of the Hilbert space of $\rho$, and $\rho$ is diagonal in
basis $T=\{\tau_{i}\}$. Then $\sigma$ is the diagonalized density
matrix of $\rho$ on basis $T=\{\tau_{i}\}$.

Now we will show $\rho$ can be diagonalized in any other set of
bases of the Hilbert space of $\rho$. If $Q_{i}$s are another set of
projective operators in this Hilbert space, and
$Q_{i}=\varsigma_{i}\varsigma^{\dag}$, $V=\{\varsigma_{i}\}$s are
also orthogonal bases of the Hilbert space of $\rho$. Similarly, we
can define matrix $Q$ constituting of bases $V=\{\varsigma_{i}\}$,
then there exists a unitary matrix $U$, with which the relationship
$P=UQ$ can be satisfied. We can rewrite density matrix $\sigma$ as
$\sigma=Q^{-1}U^{-1}\rho{U}Q$. That is, $\rho$ can be diagonalized
on basis $V=\{\varsigma_{i}\}$ after Eve operating it as
$U^{-1}\rho{U}$. That is, if a quantum state can be diagonalized on
one basis in the Hilbert space of $\rho$, it can be diagonalized on
any other bases in this Hilbert space by just rotating the states
with some unitary operation.

In DIQKD, Alice's and Bob's classical results are binary, it is
proven that their measurements can extract qubit information from
the states on the incoming particles. Then the state measured by
Alice's and Bob's devices are confined in the Hilbert space
$H_{AB}$, with dim$H_{AB}\le4$. That is, Bell basis is a set of
basis in this Hilbert space. Then if Eve can diagonalize state
$\rho$ with projective operators in the Hilbert space of $H_{AB}$,
she can diagonalize it on Bell basis.

\section{Alice and Bob's states should be identically prepared if Eve want her illegal information maximized}

This conclusion can be proven with simple mathematical technique.
Suppose there are $m$ types states prepared for Alice and Bob.

(1) When $m=1$, all states are identical.

(2) When $m=2$, they are denoted as
$\rho^{(\alpha)}_{AB}=(1-p^{(\alpha)})\rho_{|\Phi^{+}\rangle}+p^{(\alpha)}_{1}\rho_{|\Phi^{-}\rangle}+p^{(\alpha)}_{2}\rho_{|\Psi^{+}\rangle}+p^{(\alpha)}_{3}\rho_{|\Psi^{-}\rangle}$
and
$\rho^{(\beta)}_{AB}=(1-p^{(\beta)})\rho_{|\Phi^{+}\rangle}+p^{(\beta)}_{1}\rho_{|\Phi^{-}\rangle}+p^{(\beta)}_{2}\rho_{|\Psi^{+}\rangle}+p^{(\beta)}_{3}\rho_{|\Psi^{-}\rangle}$.
And their statistical representation of all systems can be written
as
$\rho_{AB}=(1-p)\rho_{|\Phi^{+}\rangle}+p_{1}\rho_{|\Phi^{-}\rangle}+p_{2}\rho_{|\Psi^{+}\rangle}+p_{3}\rho_{|\Psi^{-}\rangle}$.
These two types of states are assumed to be prepared equiprobably,
and this assumption does not loss any generality if these two types
of states can be proven to be the same.

Then Eve's information gain on these states should be less than
$\frac{N}{2}[S(\rho^{\alpha}_{AB})+S(\rho^{\beta}_{AB})]$, with $N$
is the number of the total systems shared between Alice and Bob. One
can then have
\begin{equation}
\begin{array}{l}
p_{1}^{(\alpha)}+p_{1}^{(\beta)}=2p_{1},\\
p_{2}^{(\alpha)}+p_{2}^{(\beta)}=2p_{2},\\
p_{3}^{(\alpha)}+p_{3}^{(\beta)}=2p_{3},\\
\end{array}
\end{equation}
and
\begin{equation}
\begin{array}{lll}
S_{E}&=&\frac{N}{2}[S(\rho^{\alpha}_{AB})+S(\rho^{\beta}_{AB})]\\
&=&-p_{1}^{(\alpha)}\log_{2}p_{1}^{(\alpha)}-p_{2}^{(\alpha)}\log_{2}p_{2}^{(\alpha)}-p_{3}^{(\alpha)}\log_{2}p_{3}^{(\alpha)}\\
& &-(1-p_{1}^{(\alpha)}-p_{2}^{(\alpha)}-p_{3}^{(\alpha)})\log_{2}((1-p_{1}^{(\alpha)}-p_{2}^{(\alpha)}-p_{3}^{(\alpha)}))\\
& &-p_{1}^{(\beta)}\log_{2}p_{1}^{(\beta)}-p_{2}^{(\beta)}\log_{2}p_{2}^{(\beta)}-p_{3}^{(\beta)}\log_{2}p_{3}^{(\beta)}\\
& &-(1-p_{1}^{(\beta)}-p_{2}^{(\beta)}-p_{3}^{(\beta)})\log_{2}((1-p_{1}^{(\beta)}-p_{2}^{(\beta)}-p_{3}^{(\beta)}))\\
\end{array}
\end{equation}
Substituting the relation in Eq. (7) into Eq. (8), we have
$S_{E}=S_{E}(p^{(\alpha)}_{1},p^{(\alpha)}_{2},p^{(\alpha)}_{3})$.
Varying $p^{(\alpha)}_{1}$, $p^{(\alpha)}_{2}$, and
$p^{(\alpha)}_{3}$ to make $S_{E}$ maximal, we can then have
$p^{(\alpha)}_{1}=p^{(\beta)}_{1}=p_{1}$,
$p^{(\alpha)}_{2}=p^{(\beta)}_{2}=p_{2}$, and
$p^{(\alpha)}_{3}=p^{(\beta)}_{3}=p_{3}$.

(3) Suppose all types of states are identically prepared on the $N$
systems for Eve's maximal information gain when $m=M\ge3$. That is,
Eve should make her eavesdropping optimal if $M$ types of states are
prepared on the $N$ systems shared between Alice and Bob,
$\rho^{(\alpha)}_{AB}=\rho^{(\beta)}_{AB}=\cdots=\rho^{(M)}_{AB}=\rho_{AB}$
are required, with $\rho_{AB}$ is the statistical state on the $N$
systems. That is,
$S_{E}=\frac{N}{M}[S(\rho^{(\alpha)}_{AB})+S(\rho^{(\beta)}_{AB})+\cdots+S(\rho^{(M)}_{AB})]=NS(\rho_{AB})$
calculates the optimal information obtain by Eve.

(4) When $m=M+1$, we still assume all states are prepared
equiprobably. Then
$S_{E}=\frac{N}{M+1}[S(\rho^{(\alpha)}_{AB})+S(\rho^{(\beta)}_{AB})+\cdots+S(\rho^{(M)}_{AB})+S(\rho^{(M+1)})]$.
Suppose the statistical representation for the first $M$ types of
density matrix is $\rho^{\Omega}_{AB}$,
\begin{equation}
\begin{array}{lll}
S_{E}&=&\frac{N}{M+1}[S(\rho^{(\alpha)}_{AB})+S(\rho^{(\beta)}_{AB}))+\cdots+S(\rho^{(M)}_{AB})+S(\rho^{(M+1)})]\\
&=&\frac{M}{M+1}\frac{N}{M}[S(\rho^{(\alpha)}_{AB})+S(\rho^{(\beta)}_{AB}))+\cdots+S(\rho^{(M)}_{AB})]+\frac{N}{M+1}S(\rho^{(M+1)})\\
&\le&\frac{NM}{M+1}S(\rho^{\Omega}_{AB})+\frac{N}{M+1}S(\rho^{(M+1)}).
\end{array}
\end{equation}
Denoting
$\rho^{\Omega}_{AB}=(1-p^{(\Omega)})\rho_{|\Phi^{+}\rangle}+p^{(\Omega)}_{1}\rho_{|\Phi^{-}\rangle}+p^{(\Omega)}_{2}\rho_{|\Psi^{+}\rangle}+p^{(\Omega)}_{3}\rho_{|\Psi^{-}\rangle}$,
and
$\rho^{(M+1)_{AB}}=(1-p^{(M+1)})\rho_{|\Phi^{+}\rangle}+p^{(M+1)}_{1}\rho_{|\Phi^{-}\rangle}+p^{(M+1)}_{2}\rho_{|\Psi^{+}\rangle}+p^{(M+1)}_{3}\rho_{|\Psi^{-}\rangle}$,
with similar procedure as that in step (2), we have
$p^{(\Omega)}_{1}=p^{(M+1)}_{1}=p_{1}$,
$p^{(\Omega)}_{2}=p^{(M+1)}_{2}=p_{2}$, and
$p^{(\Omega)}_{3}=p^{(M+1)}_{3}=p_{3}$. That is, the $M+1$ types of
states are also required to be identical for Eve's optimal
eavesdropping. Then all states on the $N$ systems shared between
Alice and Bob should be identically prepared if Eve want to maximize
her illegal information.

\section{Bounding Eve's eavesdropping on Alice's and Bob's measurements}

Denoting the equivalent operations in $x-z$ plane as $A_{1}$,
$A_{2}$, $B_{1}$ and $B_{2}$, their directions are assumed to be
$\theta_{1}$, $\theta_{2}$, $\varphi_{1}$ and $\varphi_{2}$
respectively. Similarly, denoting the equivalent operations in
$x^{\prime}-z^{\prime}$ plane as $A^{\prime}_{1}$, $A^{\prime}_{2}$,
$B^{\prime}_{1}$ and $B^{\prime}_{2}$, the corresponding directions
for them are $\theta^{\prime}_{1}$, $\theta^{\prime}_{2}$,
$\varphi^{\prime}_{1}$ and $\varphi^{\prime}_{2}$. The state Eve
prepares on Alice and Bob's shared systems is
$\rho_{AB}=(1-p)\rho_{|\Phi^{+}\rangle}+p_{1}\rho_{|\Phi^{-}\rangle}+p_{2}\rho_{|\Psi^{+}\rangle}+p_{3}\rho_{|\Psi^{-}\rangle}$.
Then
$S=(1-p-p_{3})[\cos(\theta_{1}-\varphi_{1})+\cos(\theta_{1}-\varphi_{2})+\cos(\theta_{2}-\varphi_{1})-\cos(\theta_{2}-\varphi_{2})]-(p_{1}-p_{2})[\cos(\theta_{1}+\varphi_{1})+\cos(\theta_{1}+\varphi_{2})+\cos(\theta_{2}+\varphi_{1})-\cos(\theta_{2}+\varphi_{2})]$
can be obtained. Furthermore, the correlation results on neighbour
bases can be depicted as
$S_{A_{1}B_{1}}=(1-p-p_{3})\cos(\theta_{1}-\varphi_{1})-(p_{1}-p_{2})\cos(\theta_{1}+\varphi_{1})$,
$S_{A_{1}B_{2}}=(1-p-p_{3})\cos(\theta_{1}-\varphi_{2})-(p_{1}-p_{2})\cos(\theta_{1}+\varphi_{2})$,
$S_{A_{2}B_{1}}=(1-p-p_{3})\cos(\theta_{2}-\varphi_{1})-(p_{1}-p_{2})\cos(\theta_{2}+\varphi_{1})$,
and
$-S_{A_{2}B_{2}}=(1-p-p_{3})\cos(\theta_{2}-\varphi_{2})-(p_{1}-p_{2})\cos(\theta_{2}+\varphi_{2})$
respectively. The symmetry of our DIQKD requires
$S_{A_{1}B_{1}}=S_{A_{1}B_{2}}=S_{A_{2}B_{1}}=-S_{A_{2}B_{2}}=\frac{S}{4}$,
then we have
\begin{equation}
\begin{array}{lll}
&
&(1-p-p_{3})\cos(\theta_{1}-\varphi_{1})-(p_{1}-p_{2})\cos(\theta_{1}+\varphi_{1}),\\
&=&(1-p-p_{3})\cos(\theta_{1}-\varphi_{2})-(p_{1}-p_{2})\cos(\theta_{1}+\varphi_{2}),\\
&=&(1-p-p_{3})\cos(\theta_{2}-\varphi_{1})-(p_{1}-p_{2})\cos(\theta_{2}+\varphi_{1}),\\
&=&-(1-p-p_{3})\cos(\theta_{2}-\varphi_{2})+(p_{1}-p_{2})\cos(\theta_{2}+\varphi_{2}).
\end{array}
\end{equation}
Based on these relationships, we can obtain
\begin{equation}
\begin{array}{lll}
(1-p-p_{3})\sin(\frac{2\theta_{1}-\varphi_{1}-\varphi_{2}}{2})\sin(\frac{\varphi_{1}-\varphi_{2}}{2})&=&-(p_{1}-p_{2})\sin(\frac{2\theta_{1}+\varphi_{1}+\varphi_{2}}{2})\sin(\frac{\varphi_{1}-\varphi_{2}}{2}),\\
(1-p-p_{3})\sin(\frac{2\varphi_{1}-\theta_{1}-\theta_{2}}{2})\sin(\frac{\theta_{1}-\theta_{2}}{2})&=&-(p_{1}-p_{2})\sin(\frac{2\varphi_{1}+\theta_{1}+\theta_{2}}{2})\sin(\frac{\theta_{1}-\theta_{2}}{2}),\\
(1-p-p_{3})\cos(\frac{2\theta_{2}-\varphi_{1}-\varphi_{2}}{2})\cos(\frac{\varphi_{1}-\varphi_{2}}{2})&=&(p_{1}-p_{2})\cos(\frac{2\theta_{2}+\varphi_{1}+\varphi_{2}}{2})\cos(\frac{\varphi_{1}-\varphi_{2}}{2}),\\
(1-p-p_{3})\cos(\frac{2\varphi_{2}-\theta_{1}-\theta_{2}}{2})\cos(\frac{\theta_{1}-\theta_{2}}{2})&=&(p_{1}-p_{2})\cos(\frac{2\varphi_{2}+\theta_{1}+\theta_{2}}{2})\cos(\frac{\theta_{1}-\theta_{2}}{2}).
\end{array}
\end{equation}

As $S$ should violate its classical bound $2$, it is reasonable to
assume that the value $p_{1}+p_{2}+p_{3}$ is small. Then
$(1-p-p_{3})$ is comparable with $1$. If
$\cos(\frac{\varphi_{1}-\varphi_{2}}{2})=0$,
$\cos(\frac{\theta_{1}-\theta_{2}}{2})=0$,
$\sin(\frac{\varphi_{1}-\varphi_{2}}{2})=0$, or
$\sin(\frac{\theta_{1}-\theta_{2}}{2})=0$, one can obtain $S\le2$,
which is not expected in DIQKD. Then Eq. (D2) can be simplified as
\begin{equation}
\begin{array}{lll}
(1-p-p_{3})\sin(\frac{2\theta_{1}-\varphi_{1}-\varphi_{2}}{2})&=&-(p_{1}-p_{2})\sin(\frac{2\theta_{1}+\varphi_{1}+\varphi_{2}}{2}),\\
(1-p-p_{3})\sin(\frac{2\varphi_{1}-\theta_{1}-\theta_{2}}{2})&=&-(p_{1}-p_{2})\sin(\frac{2\varphi_{1}+\theta_{1}+\theta_{2}}{2}),\\
(1-p-p_{3})\cos(\frac{2\theta_{2}-\varphi_{1}-\varphi_{2}}{2})&=&(p_{1}-p_{2})\cos(\frac{2\theta_{2}+\varphi_{1}+\varphi_{2}}{2}),\\
(1-p-p_{3})\cos(\frac{2\varphi_{2}-\theta_{1}-\theta_{2}}{2})&=&(p_{1}-p_{2})\cos(\frac{2\varphi_{2}+\theta_{1}+\theta_{2}}{2}).
\end{array}
\end{equation}
If both sides of Eq. (D3) are not equal to $0$, one can obtain
\begin{equation}
\begin{array}{lll}
\sin(\theta_{1}+\theta_{2})+\sin(\theta_{1}-\theta_{2})\cos(\varphi_{1}+\varphi_{2})&=&0,\\
\sin(\varphi_{1}+\varphi_{2})+\sin(\varphi_{1}-\varphi_{2})\cos(\theta_{1}+\theta_{2})&=&0.
\end{array}
\end{equation}
We can find the maximal value conditioned on the relationships in
(D4), we find $S\le2$ in this condition. By further calculation, we
can find that $\sin(\frac{2\theta_{1}+\varphi_{1}+\varphi_{2}}{2})$,
$\sin(\frac{2\varphi_{1}+\theta_{1}+\theta_{2}}{2})$,
$\cos(\frac{2\theta_{2}+\varphi_{1}+\varphi_{2}}{2})$, and
$\cos(\frac{2\varphi_{2}+\theta_{1}+\theta_{2}}{2})$ can not be
equal to $0$ at the assumption of $S>2$. Then we have $p_{1}=p_{2}$
should be satisfied. And at the same time, we have
$\theta_{2}-\theta_{1}=\frac{\pi}{2}$,
$\varphi_{2}-\varphi_{1}=-\frac{\pi}{2}$,
$\theta_{2}-\varphi_{1}=\frac{\pi}{4}$ and
$\theta_{1}-\varphi_{2}=\frac{\pi}{4}$, or
$\theta_{2}-\theta_{1}=-\frac{\pi}{2}$,
$\varphi_{2}-\varphi_{1}=\frac{\pi}{2}$,
$\theta_{2}-\varphi_{1}=-\frac{\pi}{4}$ and
$\theta_{1}-\varphi_{2}=-\frac{\pi}{4}$.


\begin{thebibliography}{99}
\bibitem{bb84}
C. H. Bennett and G. Brassard, \emph{in Proceedings of the IEEE
International Conference on Computers, Systems and Signal
Processing}, Bangalore, India (IEEE, New York), PP. 175 (1984).

\bibitem{epr91}
A. K. Ekert, \emph{Phys. Rev. Lett.} \textbf{67}, 661 (1991).

\bibitem{gsin02}
N. Gisin, G. Ribordy, W. Tittel, and H. Zbinden, \emph{Rev. Mod.
Phys.} \textbf{74}, 145 (2002).

\bibitem{lc99}
H.-K. Lo and H. F. Chau, \emph{Science} \textbf{283}, 2050¨C2056
(1999).

\bibitem{sp00}
P. W. Shor and J. Preskill, \emph{Phys. Rev. Lett.} \textbf{85}, 441
(2000).

\bibitem{higm95}
B. Huttner, N. Imoto, N. Gisin and T. Mor, Phys. Rev. A \textbf{51},
1863 (1995).

\bibitem{qflm07}
B. Qi, C.-H. F. Fung, H.-K. Lo and X. Ma, Quant. Inf. Comp. 7, pp.
73-82 (2007).

\bibitem{zfqcl08}
Y. Zhao, C.-H. F. Fung, B. Qi, C. Chen and H.-K. Lo, Phys. Rev. A
78, 042333 (2008).

\bibitem{fqtl07}
C.-H. F. Fung, B. Qi, K. Tamaki and H.-K. Lo, Phys. Rev. A 75,
032314 (2007).

\bibitem{xql10}
F. Xu, B. Qi and H.-K. Lo, New J. Phys. 12, 113026 (2010).

\bibitem{lwwesm10}
L. Lydersen, C. Wiechers, C. Wittmann, D. Elser, J. Skaar and V.
Makarov, Nature Photonics 4, pp. 686-689 (2010); Z. L. Yuan, J. F.
Dynes and A. J. Shields, Nature Photonics 4, pp. 800-801 (2010); L.
Lydersen, C. Wiechers, C. Wittmann, D. Elser, J. Skaar and V.
Makarov, Nature Photonics 4, 801 (2010).

\bibitem{gllskm11}
I. Gerhardt, Q. Liu, A. Lamas-Linares, J. Skaar, C. Kurtsiefer and
V. Makarov, Nature Comm. 2, 349 (2011).

\bibitem{my98}
D. Mayers and A. C.-C. Yao, in Proceedings of the 39th Annual
Symposium on Foundations of Computer Science (FOCS98), (IEEE
Computer Society, Washington, DC, 1998), p. 503.

\bibitem{abgmps07}
A. Ac\'{\i}n, N. Brunner, N. Gisin, S. Massar, S. Pironio and V.
Scarani, \emph{Phys. Rev. Lett.} \textbf{98}, 230501 (2007);

\bibitem{pabgms09}
S. Pironio, A. Ac\'{\i}n, N. Brunner, N. Gisin, S. Massar and V.
Scarani, \emph{New J. Phys.} \textbf{11}, 045021 (2009).

\bibitem{chsh69}
J. Clauser \emph{et. al.}, \emph{Phys. Rev. Lett.} \textbf{23}, 880
(1969).

\bibitem{bell65}
J. S. Bell, \emph{Phsycis} \textbf{1}, 195 (1965).

\bibitem{ma07}
X. Ma , C.-H. F. Fung , and H.-K. Lo, \emph{Phys. Rev . A}
\textbf{76} 012307, (2007).

\bibitem{renner05a}
R. Renner, N. Gisin, B. Kraus, \emph{Phys. Rev. A} \textbf{72},
012332 (2005).

\bibitem{renner05}
R. Renner, Ph.D. thesis, ETH No. 16242, arXiv:quant-ph/ 0512258.

\bibitem{renner07}
R. Renner, \emph{Nature Phys.} \textbf{3}, 645 (2007).

\bibitem{holevo73}
A. S. Holevo, \emph{Probl. Peredachi. Inf.}, \textbf{ 9}, 3 (1973).

\bibitem{cabello00}
A. Cabello, \emph{Phys. Rev. Lett.}, \textbf{85}, 5635 (2000).

\bibitem{nielsen00}
M. A. Nielsen, and I. L. Chuang, \emph{Quantum computation and
quantum information.} Cambridge University Press, (2000).

\bibitem{tsirelson80}
B. S. Tsirelson . \emph{Lett. Math. Phys.}, \textbf{4}, 93 (1980).

\bibitem{tsirelson93}
B. Tsirelson, \emph{Hadronic Journal Supplement}, \textbf{8}, 329
(1993).

\bibitem{masanes06}
L. Masanes, \emph{Phys. Rev . Lett.}, \textbf{97}, 050503 (2006).

\bibitem{brub99}
D. Bru{\ss}, \emph{Phys. Rev. A}, \textbf{60}, 4344 (1999).

\bibitem{ckw00}
V. Coffman, J. Kundu, and W. K. Wootters, \emph{Phys. Rev. A},
\textbf{61} 052306 (2000).

\bibitem{kw04}
M. Koashi, and A. Winter, \emph{Phys. Rev. A} \textbf{69}, 022309
(2004).

\bibitem{eberhard93}
P. H. Eberhard, \emph{Phys. Rev. A}, \textbf{47}, R747 (1993).

\bibitem{brunner07}
N. Brunner, N. Gisin, V. Scarani, and C. Simon, \emph{Phys. Rev.
Lett.}, \textbf{98}, 220403 (2007).

\bibitem{cabello07}
A. Cabello and J.-A. Larsson, \emph{Phys. Rev. Lett.} \textbf{98},
220402 (2007).

\bibitem{biham97}
E. Biham and T. Mor, \emph{Phys. Rev. Lett.}, \textbf{79}, 4034
(1997).

\bibitem{biham97a}
E. Biham and T. Mor. \emph{Phys. Rev. Lett.}, \textbf{78}, 2256
(1997).

\bibitem{gerhardt11}
I. Gerhardt, Q. Liu, A. Lamas-Linares, J. Skaar, V. Scarani, V.
Makarov, and C. Kurtsiefer, \emph{Phys. Rev. Lett.}, \textbf{107},
170404 (2011).

\bibitem{gisin10}
N. Gisin, S. Pironio, and N. Sangouard, \emph{Phys. Rev. Lett.},
\textbf{105}, 070501 (2010).

\end{thebibliography}
\end{document}